\documentclass{pasj00}
\draft

\begin{document}
\SetRunningHead{Tatematsu et al.}{Chemical Variation in Orion}
\Received{}
\Accepted{}

\title{Chemical Variation in Molecular Cloud Cores in the Orion A Cloud}

\author{Ken'ichi \textsc{Tatematsu},\altaffilmark{1,2}
Tomoya \textsc{Hirota},\altaffilmark{1,2}
Ryo \textsc{Kandori},\altaffilmark{1}
and Tomofumi \textsc{Umemoto}\altaffilmark{1}}
\altaffiltext{1}{National Astronomical Observatory of Japan, 
2-21-1 Osawa, Mitaka, Tokyo 181-8588}
\altaffiltext{2}{Department of Astronomical Science, 
Graduate University for Advanced Studies, 
2-21-1 Osawa, Mitaka, Tokyo 181-8588}
\email{k.tatematsu@nao.ac.jp, tomoya.hirota@nao.ac.jp,
r.kandori@nao.ac.jp, umemoto.tomofumi@nao.ac.jp}


%

\KeyWords{ISM: clouds
---ISM: individual (Orion Nebula, Orion Molecular Cloud)
---ISM: molecules
---ISM: structure---stars: formation} 

\maketitle

\begin{abstract}
We have observed molecular cloud cores 
in the Orion A giant molecular cloud (GMC)
in CCS, HC$_3$N, DNC, and HN$^{13}$C to study
their chemical characteristics.
We have detected CCS in the Orion A GMC for the first time.
CCS was detected in about a third of the observed cores.
The cores detected in CCS are not localized
but are widely distributed over the Orion A GMC. 
The CCS peak intensity of the core tends to be
high in the southern region of the Orion A GMC.
The HC$_3$N peak intensity of the core also tends to be high
in the southern region, 
while there are HC$_3$N intense cores
near Orion KL, which is not seen in CCS.  
The core associated with Orion KL shows
broad HC$_3$N line profile, and star formation activity near Orion KL 
seems to enhance the HC$_3$N emission.
The column density ratio of NH$_3$ to CCS is lower near the 
middle of the filament, and is higher toward the northern and southern
regions along the Orion A GMC filament.
This ratio is known to trace the chemical evolution in nearby
dark cloud cores, but seems to be affected by core gas temperature 
in the Orion A GMC:
cores with low NH$_3$ to CCS column density ratios tend to
have warmer gas temperature.
The value of the column density ratio of DNC to HN$^{13}$C
is generally similar to that in dark cloud cores,
but becomes lower around Orion KL due to
higher gas temperature.

\end{abstract}

\section{Introduction}

Most of stars in the Galaxy form in giant molecular clouds (GMCs).
Molecular cloud cores are known to be the sites of star formation
(e.g., \cite{bei86}).
The evolution of molecular cloud cores 
in GMCs is less understood,
compared with that of nearby dark cloud cores.
In nearby dark clouds, molecules such as CCS, HC$_3$N, NH$_3$, and N$_2$H$^+$,
and the neutral carbon atom C$^0$
are known to be good tracers of chemical evolution
(e.g., \cite{hir92,suz92,ben98,mae99,hir02,hir09}).
The carbon-chain molecules, CCS and HC$_3$N tend to trace 
the early chemical 
evolutionary stage, whereas N-bearing molecules, 
NH$_3$ and N$_2$H$^+$ tend to trace
the late stage.
We wonder how different the chemical properties of molecular cloud cores
in GMCs are, compared with those in nearby dark clouds.

The Orion A cloud is an archetypal 
GMC.
This cloud has been extensively mapped in $^{12}$CO $J$ = 1$-$0
\citep{hey92}, 
in $^{13}$CO $J$ = 1$-$0 \citep{bal87,nag98}, 
in CS $J$ = 1$-$0 \citep{tat93},
in CS $J$ = 2$-$1 \citep{tat98}, 
in NH$_3$ \citep{bat83,ces94}, 
in H$^{13}$CO$^+$ $J$ = 1$-$0 \citep{ike07}
and
in N$_2$H$^+$ $J$ = 1$-$0 \citep{tat08}.
The dust continuum emission in the Orion A cloud was studied by
\citet{chi97}, \citet{lis98}, and \citet{joh99}.

The chemical variation has been suggested in limited regions in the 
Orion A GMC.
\citet{ung97} studied the region near Orion KL in many molecular lines, 
and have shown difference in molecular abundance between 
the Orion KL region, 
the Orion Bar, and the molecular ridge.
\citet{tat93} studied the region covering
Orion KL and OMC-2 (see their Figure 4),
and found that NH$_3$ tends to be stronger in the north
(OMC-2 region) 
while
CS tends to be stronger in the south.
\citet{tat08} showed that
N$_2$H$^+$ is widely distributed over the $\int$-shaped filament of
the Orion A GMC, and
pointed out that NH$_3$ and N$_2$H$^+$ is relatively
strong toward OMC-2, compared with the H$^{13}$CO$^+$ and CS distribution.
\citet{joh03} astrochemically compared submillimeter dust continuum sources
in the Orion A GMC.

In this study, we investigate the global chemical characteristics of
molecular cloud cores in the Orion A GMC
through new molecular-line
observations.
The purposes of this study are 
(1) to search for the CCS emission, which may represent 
the young molecular cloud material,
(2) to study the chemical properties of the cloud cores through 
the line intensity variation and the column density ratio variation
ALONG the Orion A GMC filament,
and 
(3) to compare the chemical properties of Orion cores with those of 
dark cloud cores.

The distance to the Orion A cloud is estimated to be 418 pc 
\citep{kim08}. At this distance,
1$\arcmin$ corresponds to 0.12 pc.

\newpage

\section{Observations}

Observations were carried out by using the 45 m radio telescope
of Nobeyama Radio Observatory\footnote{Nobeyama Radio Observatory 
is a branch of the National Astronomical Observatory of Japan, 
National Institutes of Natural Sciences.} from 2009 April 3 to 6.  
The employed receiver front ends were
the single-element SIS receivers, ``T100V'' and ``S40''.
We observed 
HN$^{13}$C $J$ = 1$-$0 at 87.090859 GHz and
DNC $J$ = 1$-$0 at 76.305717 GHz
by using the 2SB receiver ``T100V'', and
CCS $J_N$ = 4$_3-3_2$ at 45.379033 GHz \citep{yam90}
and 
HC$_3$N $J$ = 5$-$4 at 45.490316 GHz 
(the frequencies of the hyperfine-components are 
averaged with weighting) \citep{lov04}
by using the SSB receiver ``S40''.
All the lines were observed simultaneously, by using 
two receivers with a polarization
splitter.
The half-power beamwidth of the telescope
with the T100V and S40 receivers were 18$\farcs$4$\pm$0$\farcs$1 
and 38$\farcs$5$\pm$0$\farcs$1 at 86 and 43 GHz, respectively.
The receiver back end was acousto-optical spectrometers.
The spectral resolution 
was 37 kHz  (corresponding to $\sim$ 0.25 km s$^{-1}$ at 45 GHz,
$\sim$ 0.15 km s$^{-1}$ at 76 GHz, and
0.13 km s$^{-1}$ at 87 GHz).
Spectra were obtained in the position-switching mode.
The employed off position was either ($\Delta$ R. A., $\Delta$ Dec.)
= ($-$30$\arcmin$, 5$\arcmin$), (0$\arcmin$, $-$100$\arcmin$),
or (107$\arcmin$, $-$95$\arcmin$)
with respect to Orion KL.
The intensity is reported in terms of the corrected
antenna temperature $T_A^*$.
The main-beam efficiency 
was 0.42$\pm$0.02 for ``T100V'', 
and  0.73$\pm$0.04 for ``S40''.
The telescope pointing was established by observing Orion KL
in the 43-GHz SiO maser line every 60$-$80 min.

The positions for the observations were chosen from intensity 
peak positions of 125 CS
cores in \citet{tat93}.
The core name used here is `TUKH' followed by three digits
of the core number of \citet{tat93}.
Note that the degrees of the declination were printed incorrectly for cores 
TUKH068 and TUKH070 in Table 1 of \citet{tat93}: ``$-$5$\degree$'' 
should read ``$-$6$\degree$.''
We selected 81 peak positions toward which 
both CS $J$ = 1$-$0 and CS $J$ = 2$-$1
were clearly and consistently detected without velocity-component confusion,
on the basis of \citet{tat98}.
In the CCS and HC$_3$N observations, we lost some data due to 
a mistake in the receiver set-up,
and we do not use data with high noise levels .
The final number of the cores successfully observed in 
CCS and HC$_3$N is 62 (out of 81).
The observed data were reduced by using the software package ``NewStar''
of Nobeyama Radio Observatory.

\section{Results and Discussion}

\subsection{Line Intensities and Their Variations} 

\begin{figure}
  \begin{center}
    \FigureFile(150mm,150mm){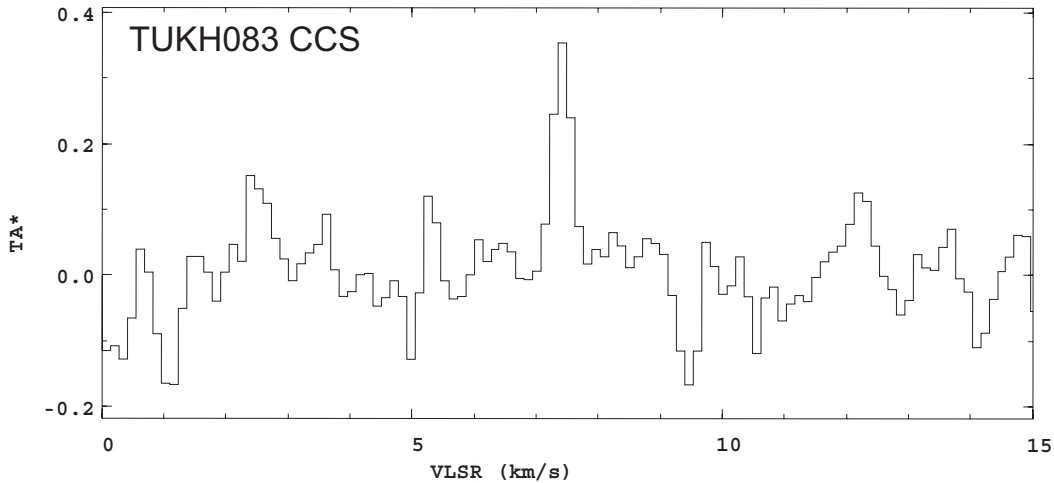}
  \end{center}
  \caption{CCS $J_N$ = 4$_3$-3$_2$ spectrum observed toward TUKH083.

}\label{fig:figure1}
\end{figure}

\begin{figure}
  \begin{center}
    \FigureFile(150mm,150mm){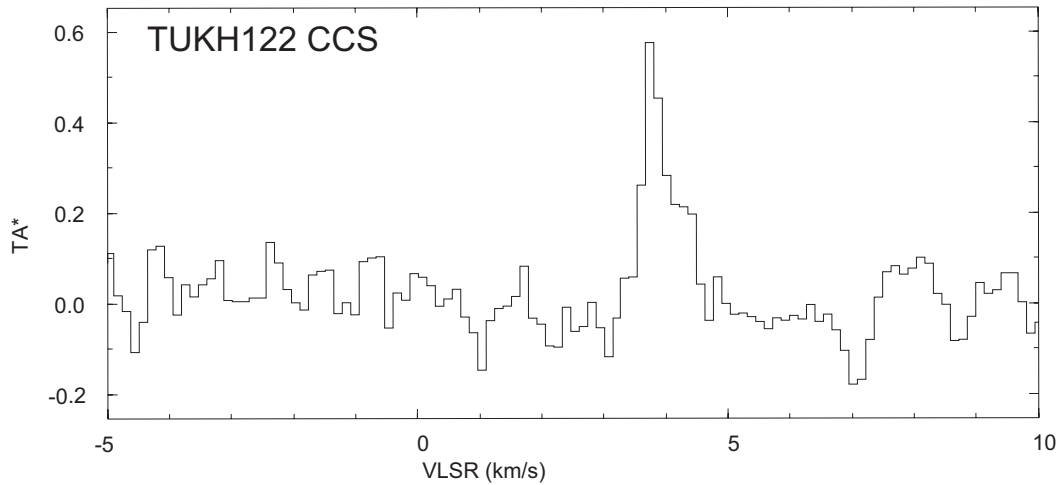}
  \end{center}
  \caption{CCS $J_N$ = 4$_3$-3$_2$ spectrum observed toward TUKH122.

}\label{fig:figure2}
\end{figure}

Table 1 summarizes the observed line parameters:
the peak intensity $T_A^*$, the LSR velocity $\Delta v_{LSR}$,
and the linewidth of CCS, HC$_3$N. HN$^{13}$C, and DNC.
The HC$_3$N peak intensity in the table represents $T_A^*$ 
of the observed spectrum
rather than the hyperfine-component intensity.
`Three dots' in $T_A^*$ means either `not observed' or 
`omitted because noise level is too high 
(the rms noise level is larger than
0.14 K).'
Figures \ref{fig:figure1} and \ref{fig:figure2} show the CCS spectra obtained toward
cores TUKH083 and TUKH122, respectively.
To our knowledge, the detection of CCS has not been
reported in Orion GMC cores previously, and our observations
serve as the first CCS detection in this cloud.  
Because it is suggested that
CCS traces 
young molecular gas ($<$ 10$^6$ yr; e.g. \cite{suz92}) in
nearby dark clouds, 
it is very interesting that the Orion A GMC, whose age is at 
least of order 10$^7$ yr (e.g. Bally et al. 1987), harbors an early-type 
molecule CCS. 
Furthermore, the detection rate is high.
We detected CCS in 20 core peaks out of 62, implying the detection
rate is 32\%.  
For HC$_3$N, the detection rate is 53\% (33 out of 62).
The detection rates for DNC and HN$^{13}$C are 
32\% (26 out of 81) and 37\% (30 out of 81), respectively.
\citet{tat08} did not detected CCS anywhere 
in the OMC-2/3 region, but better rms noise levels
in the present study allowed us to detect
one core TUKH003 in CCS out
of six observed cores in the OMC-2/3 region, resulting in a 
detection rate of 17\% in this region.
We compare our detection rates with previous studies.
\citet{hir09} observed CCS and HC$_3$N toward nearby dark cloud cores
and their detection rates were 42\% and 42\%,
and \citet{sak08} observed the same molecules 
toward infrared dark clouds and
their detection rates were 0\% and 78\%,
respectively.
These studies including the present study used the same telescope.
The rms noise level was $\sim$0.1 K (CCS) and $\sim$0.15 K (HC$_3$N)
in the present study,  $\sim$0.05 K (CCS) and $\sim$0.1 K (HC$_3$N)
in \cite{hir09}, and  $\sim$0.2 K in \citet{sak08}.
The non-detection of CCS in infrared dark clouds 
is in contrast with the detection 
rate of CCS in Orion A GMC cores and in nearby dark cloud cores.
This can be partly due to somewhat higher rms noise level for observations
toward infrared dark clouds.
Infrared dark clouds are farther than Orion A GMC and nearby dark clouds,
and smaller beam filling factor of CCS may explain difference.
However, the HC$_3$N detection rate is higher in infrared dark clouds,
and it is impossible to explain difference between 
CCS and HC$_3$N detection rates
only in terms of sensitivity and beam dilution.
It is most likely that different detection rates
reflect difference in chemical properties between 
infrared dark clouds and the others.
The detection rates of CCS and HC$_3$N in Orion are not very different
from those in nearby dark clouds.
It seems that 
Orion GMC cores are more similar to nearby dark cloud cores
than infrared dark clouds.

\begin{figure}
  \begin{center}
    \FigureFile(150mm,150mm){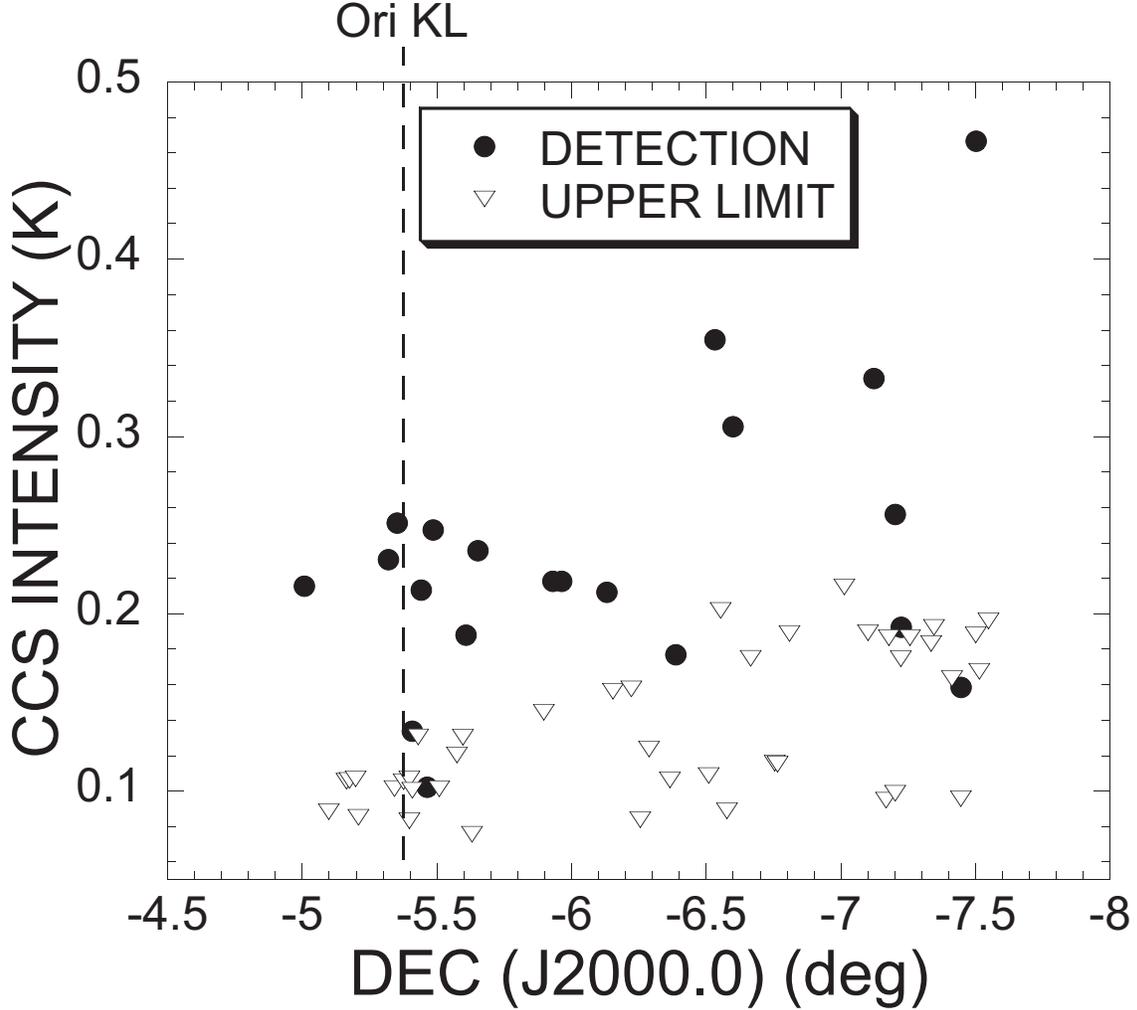}
  \end{center}
  \caption{The CCS $J_N$ = 4$_3$-3$_2$ peak intensity $T_A^*$ 
against the declination.
}\label{fig:figure3}
\end{figure}

\begin{figure}
  \begin{center}
    \FigureFile(150mm,150mm){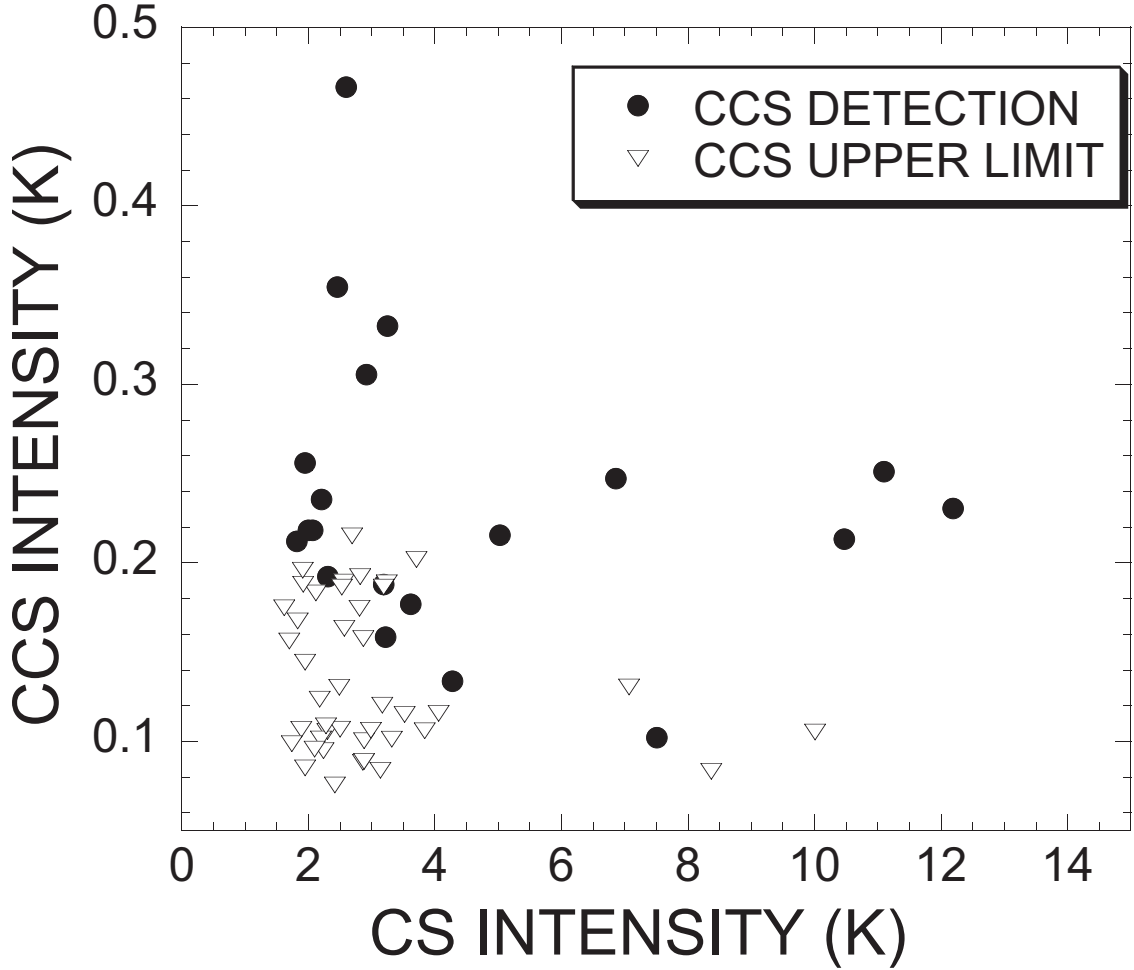}
  \end{center}
  \caption{The CCS $J_N$ = 4$_3$-3$_2$ intensity $T_A^*$ toward Orion cores
is plotted against the CS $J$ = 1$\rightarrow$0 
intensity from \citet{tat93}.

}\label{fig:figure4}
\end{figure}

\begin{figure}
  \begin{center}
    \FigureFile(150mm,150mm){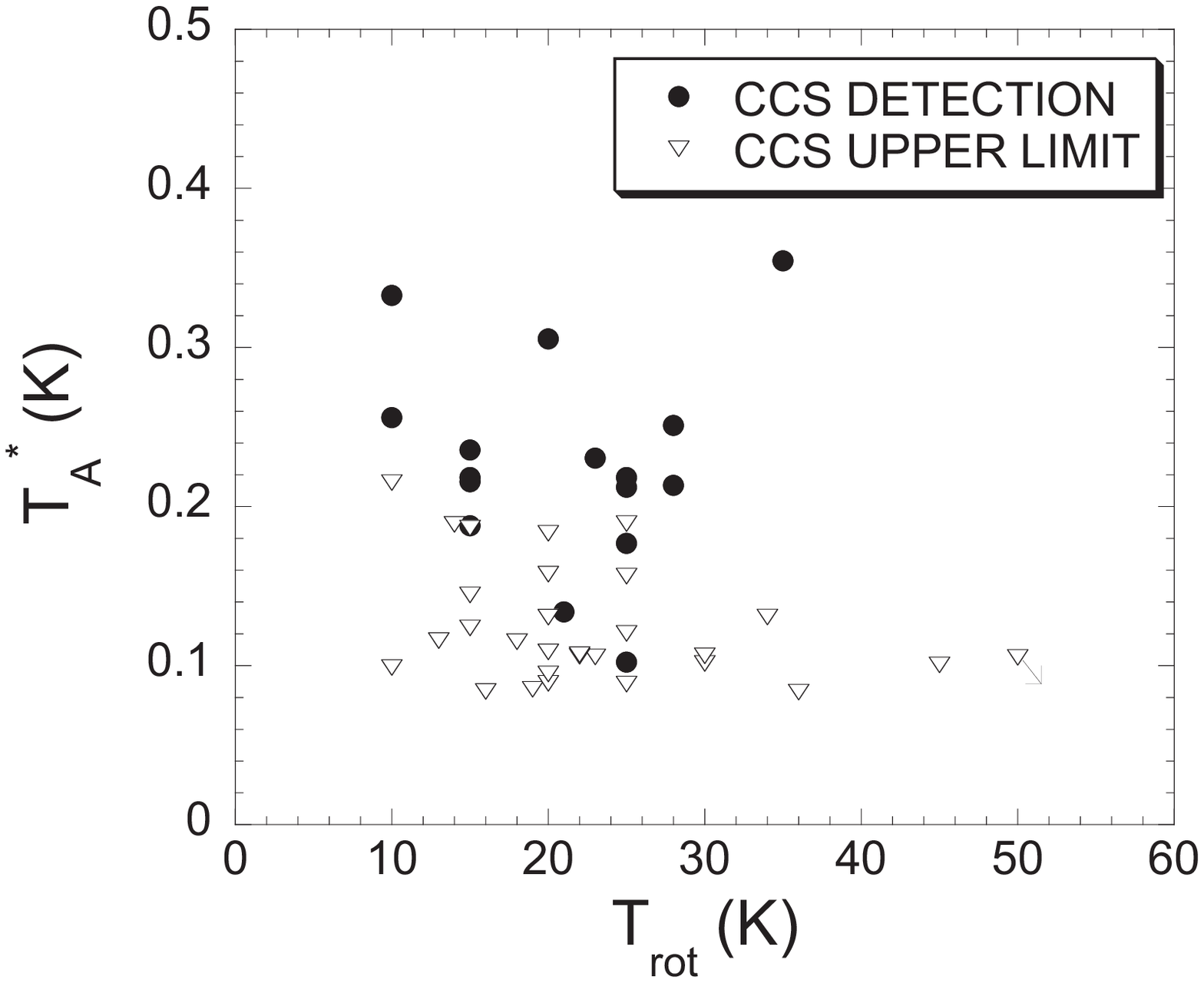}
  \end{center}
  \caption{CCS $J_N$ = 4$_3$-3$_2$ intensity $T_A^*$ toward Orion cores
is plotted against NH$_3$ rotation temperature $T_{rot}$ \citep{wil99}.
Orion KL is shown as the upper limit to the CCS intensity
and the lower limit to the NH$_3$ rotation temperature.
The most intense CCS core in Figure \ref{fig:figure3} is not plotted here
because the $T_{rot}$ is not listed.
}\label{fig:figure5}
\end{figure}

The observed lines are thought to be optically thin.
We investigate the peak intensity variation
along the Orion A GMC filament.
Figure \ref{fig:figure3} plots the CCS peak 
intensity $T_A^*$ against the declination.
Note that the Orion A GMC is roughly elongated in the NS direction
(e.g., \cite{bal87}).
CCS-detected cores are not localized,
but are distributed over the Orion A GMC.
It seems that the CCS peak 
intensity is stronger in the south (Dec. $<$ $-$6.5 deg).
We wonder whether there is any relationship between the CS and CCS intensities,
because these molecules are thought to be chemically related \citep{suz92}.
Figure \ref{fig:figure4} compares the CCS peak 
intensity with the CS peak intensity of \citet{tat93}.
There is no positive correlation between these intensities.
Most intense CCS cores are weak in CS.
From the CS intensity, it is hard to predict the CCS detection.
Figure \ref{fig:figure5} compares the CCS peak intensity with the 
NH$_3$ rotation temperature \citep{wil99}.
CCS was not detected toward Orion KL, which is warm ($T_{rot}$ $>$ 50 K and
kinetic temperature $T_{kin}$ = 70$-$150 K; \cite{gen89}; \cite{wil99}).  
See \citet{dan88} for the $T_{rot}-T_{kin}$ relation.

\begin{figure}
  \begin{center}
    \FigureFile(150mm,150mm){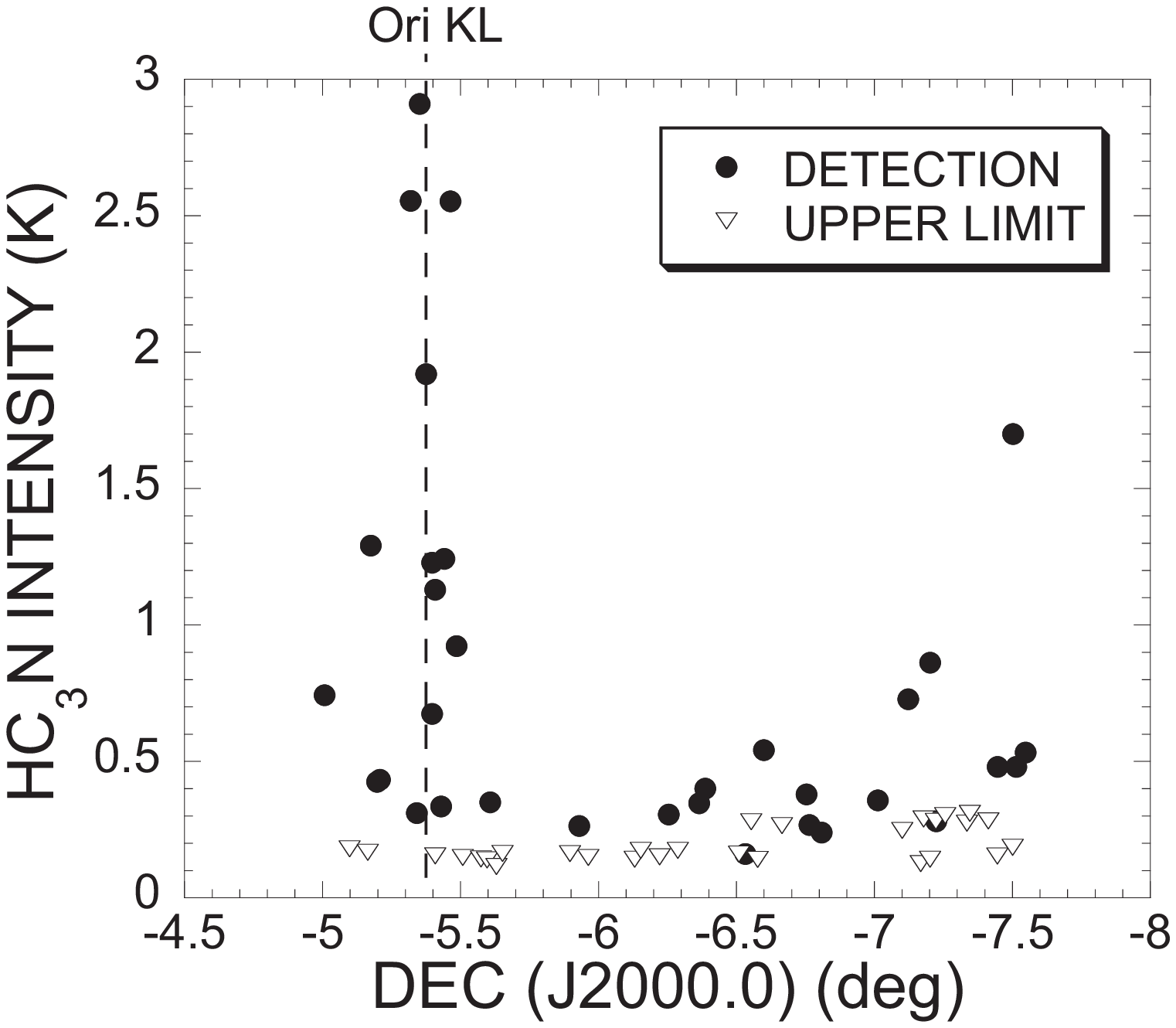}
  \end{center}
  \caption{The HC$_3$N peak intensity $T_A^*$ against the declination.
}\label{fig:figure6}
\end{figure}

Next, we show an observational result in  HC$_3$N. 
Figure \ref{fig:figure6} plots the HC$_3$N peak 
intensity against the declination.
Again, 
HC$_3$N-detected cores are not localized, but distributed over the Orion A 
GMC filament.
There are two characteristics in the HC$_3$N peak intensity distribution.
First, the HC$_3$N peak intensity tends to be stronger toward the south,
as the CCS peak intensity distribution does.
Because CCS and HC$_3$N are known to be early-type molecules in nearby
dark clouds,
this similarity can be understood as they trace similar 
chemical evolutionary
stages.
Second, the HC$_3$N intensity is very strong around 
Orion KL (Dec. = $-$5.375 deg), which is
not seen in the CCS intensity distribution.
The HC$_3$N linewidth toward Orion KL is broad ($\sim$ 6 km s$^{-1}$), and
it seems that the HC$_3$N emission is also affected by star formation 
activities such as molecular outflows.

\begin{figure}
  \begin{center}
    \FigureFile(150mm,150mm){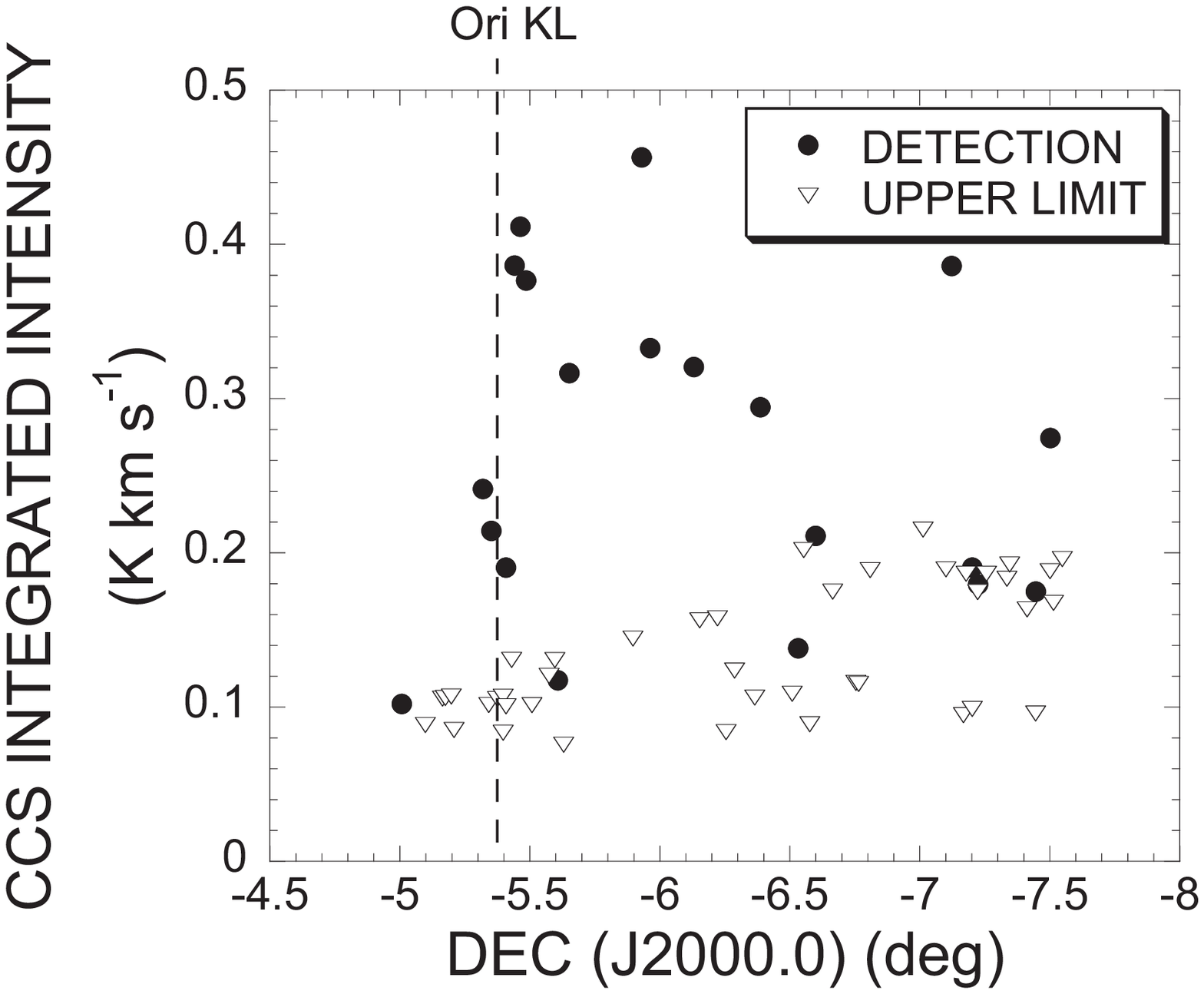}
  \end{center}
  \caption{The CCS $J_N$ = 4$_3$-3$_2$ integrated intensity against 
the declination.

}\label{fig:figure7}
\end{figure}

\begin{figure}
  \begin{center}
    \FigureFile(150mm,150mm){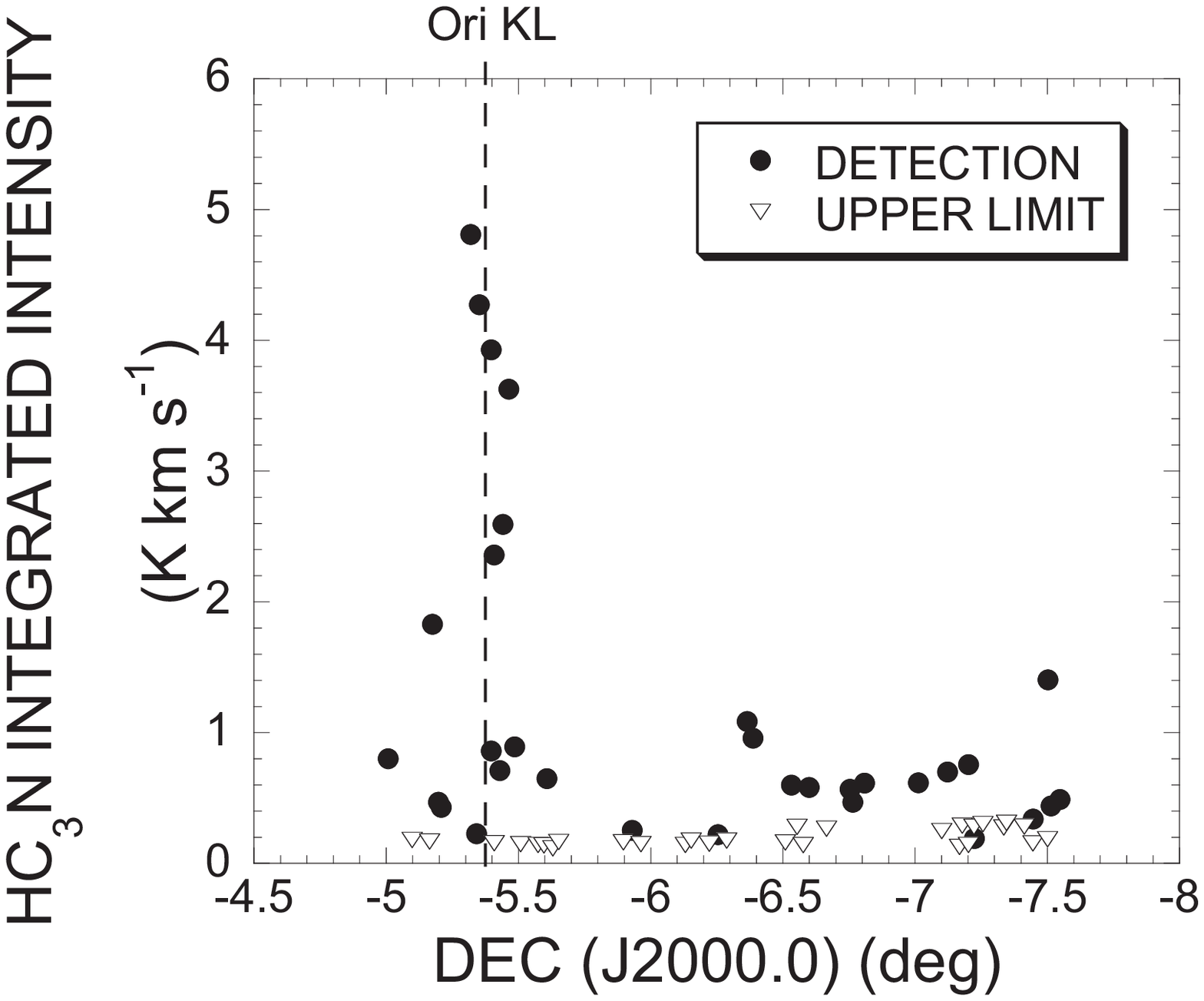}
  \end{center}
  \caption{The HC$_3$N integrated intensity against 
the declination.  The most intense core associated with Orion KL,
which has $\int T_A^* dv$ = 12.8 K km s$^{-1}$, is omitted in this figure
}\label{fig:figure8}
\end{figure}

We investigate the variation also in the velocity-integrated intensity 
$\int~T_A^*~dv$.
Figures \ref{fig:figure7} and \ref{fig:figure8} show the
integrated intensities of CCS and HC$_3$N against the declination.
The integrated intensity is calculated for the best-fit Gaussian 
to the observed 
spectrum.
The CCS and HC$_3$N linewidths are larger near the Orion Nebula region,
and become smaller toward the south (not shown) as observed in other moleculs
\citep{tat93}.
The integrated intensity does not tend to be high toward the south
in CCS, and tends to be high only weakly toward the south in HC$_3$N.
It seems that the variation in the peak and integrated intensities
are similar in CCS and HC$_3$N except for the strong HC$_3$N emission
near Orion KL.

\subsection{Column Densities} 

We investigate the chemical properties of the cloud cores by
using the column density ratio of the molecules.
The column density is calculated by assuming 
local thermodynamic equilibrium (LTE)
with two excitation temperatures:
$T_{ex}$ = $T_{rot}$ and $T_{ex}$ = $T_{rot}$/2.
$T_{rot}$ is taken from NH$_3$ observations of \citet{wil99}.
The NH$_3$ rotation temperature is very close to
the gas kinetic temperature T$_k$ for T$_k$ $\lesssim$ 30 K 
\citep{dan88}.
We use $T_{ex}$ = $T_{rot}$ as (almost) thermalized cases
and $T_{ex}$ = $T_{rot}$/2 as subthermal cases.
When \citet{wil99} lists an upper or lower limit to the rotation
temperature, we employ the limit value as $T_{rot}$.
When \citet{wil99} does not list a rotation temperature due to no detection,
we assume $T_{rot}$ = 20 K, i.e., $T_{ex}$ = 20 and 10 K.
The dipole moments of CCS, HC$_3$N, DNC, and HN$^{13}$C are
2.81 \citep{mur90}, and 3.724, 3.050, and 2.699 Debye 
(JPL Catalog\footnote{JPL Catalog is available 
at http://spec.jpl.nasa.gov/ftp/pub/catalog/catform}), 
respectively.
The upper limit to the column density is defined as 3$\sigma$
for 1 km s$^{-1}$ velocity width.

\begin{figure}
  \begin{center}
    \FigureFile(150mm,150mm){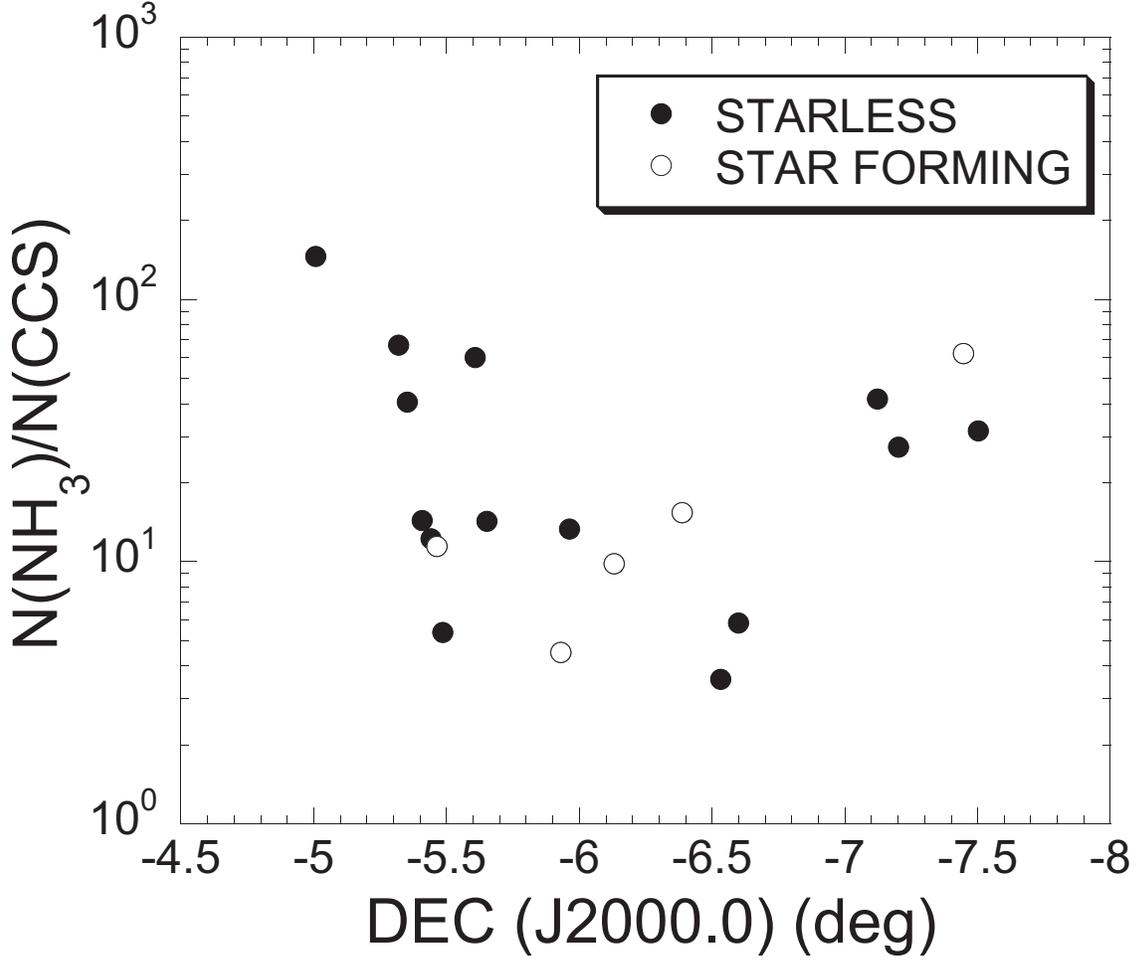}
  \end{center}
  \caption{$N$(NH$_3$)/$N$(CCS) column density ratio 
is plotted against the declination.
The CCS excitation temperature is assumed to be equal to $T_{rot}$.

}\label{fig:figure9}
\end{figure}

\begin{figure}
  \begin{center}
    \FigureFile(150mm,150mm){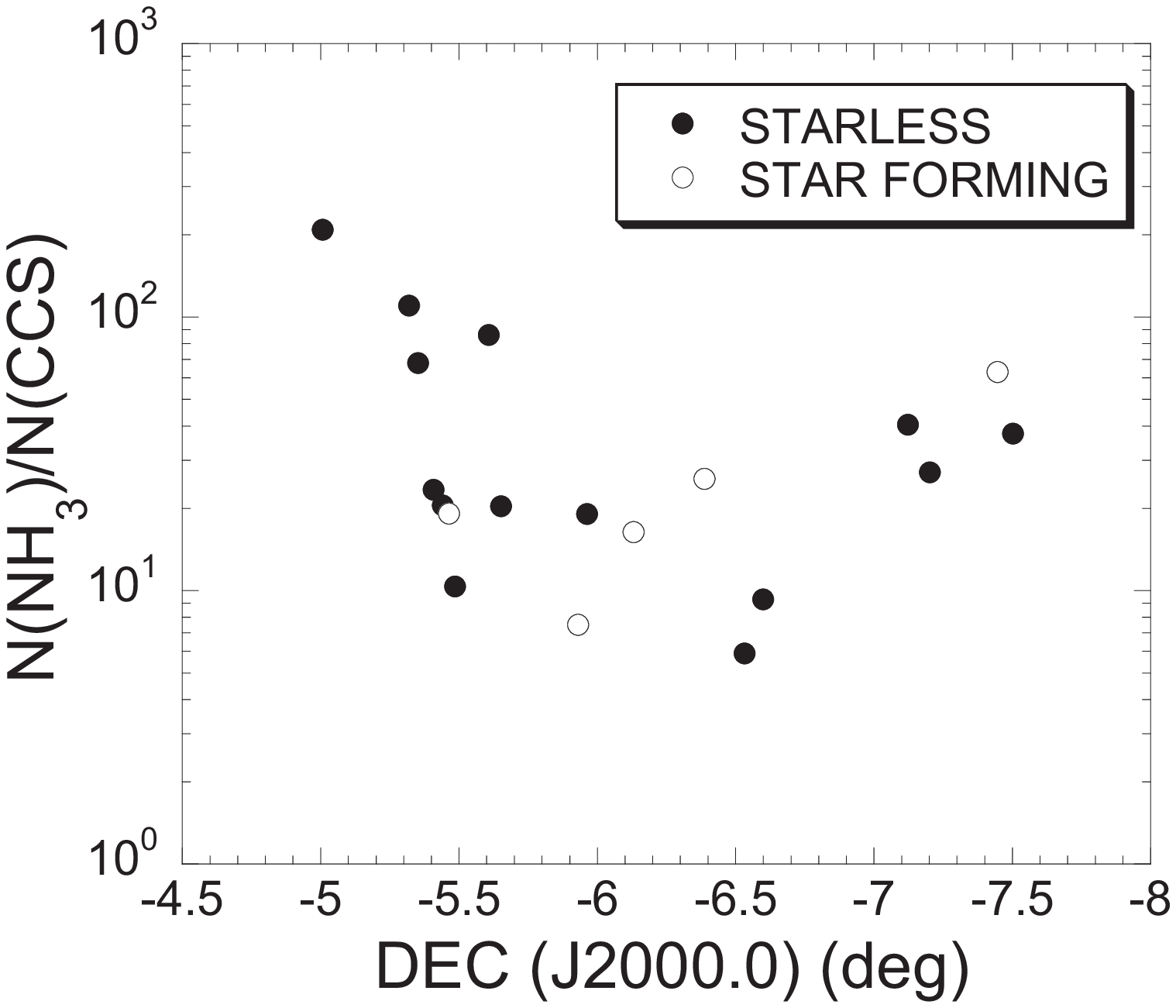}
  \end{center}
  \caption{The same as Figure \ref{fig:figure9} 
but for the CCS excitation temperature equal to $T_{rot}$/2.
}\label{fig:figure10}
\end{figure}

To see the chemical properties,
we compare the column density of the late-type molecule NH$_3$ 
to the early-type molecule CCS.
Figures \ref{fig:figure9} and \ref{fig:figure10} plot the 
$N$(NH$_3$)/$N$(CCS) 
column density ratio against the declination, with CCS
excitation temperatures of $T_{rot}$ and $T_{rot}$/2, respectively.
The ratio is not constant, is lowest in the middle, 
and becomes higher toward the north (Dec. $\gtrsim$ $-$5.5 deg) 
and south (Dec. $<$ $-$7 deg).
This suggests that the Orion A GMC filament has global chemical variation
along the filament.
We found that 
the CCS and HC$_3$N peak 
intensities are stronger in the south region.
However, this tendency is not seen (CCS) or not very prominent (HC$_3$N) 
in the integrated intensity.
Furthermore, the NH$_3$ column density is also higher in the south.
As a result, southern cores are found to have large $N$(NH$_3$)/$N$(CCS).
The logarithmic column density ratio of 
NH$_3$ to CCS, log $N$(NH$_3$)/$N$(CCS) ranges from 0.9 to 2.3.
If CCS and NH$_3$ represent early and late evolutionary stages, respectively,
also in GMCs,
the Orion A GMC is chemically younger in the middle
and older in the north and south regions.

We plot starless cores and star-forming cores separately
in Figures \ref{fig:figure9} and \ref{fig:figure10} (and hereafter).
Here, the cores associated with molecular outflows and/or IRAS point sources
are classifies as star-forming cores (Table 1 of \cite{tat93}).
Furthermore, we visually investigated the Spitzer 24 $\mu$m MIPS images. 
We found that observed cores TUKH008, TUKH056, TUKH079, 
TUKH087, TUKH092, TUKH097, 
TUKH111, and TUKH125 are associated with
point sources. 
Cores TUKH092 and TUKH097 are
associated with several and two point sources, respectively,
while the other cores are associated with one point source each.
These are also classified as star-forming cores.
On the other hand, core TUKH89 associated with Morgan 18 
does not accompany a Spitzer source.
\citet{mor91} identified a molecular outflow but
its distribution is not well defined.
We exclude this core as a star-forming core and classified 
as a starless core.
Note that identification of the Spitzer MIPS images is 
difficult near the Orion Nebula due to intense 
diffuse infrared emission.
We investigate the relationship between $N$(NH$_3$)/$N$(CCS)
and the association with young stellar objects.
High $N$(NH$_3$)/$N$(CCS) cores ($N$(NH$_3$)/$N$(CCS) $\gtsim$ 30) 
are mostly starless cores,
which is in contrast with results in dark cloud cores 
(\cite{suz92}; \cite{hir09}).
The reason for this is not clear.
The high $N$(NH$_3$)/$N$(CCS) 
cores are located in outer regions w.r.t. the central part
of the Orion Nebula, and therefore the poor MIPS identification
near the Orion Nebula cannot be the cause of this.

\begin{figure}
  \begin{center}
    \FigureFile(150mm,150mm){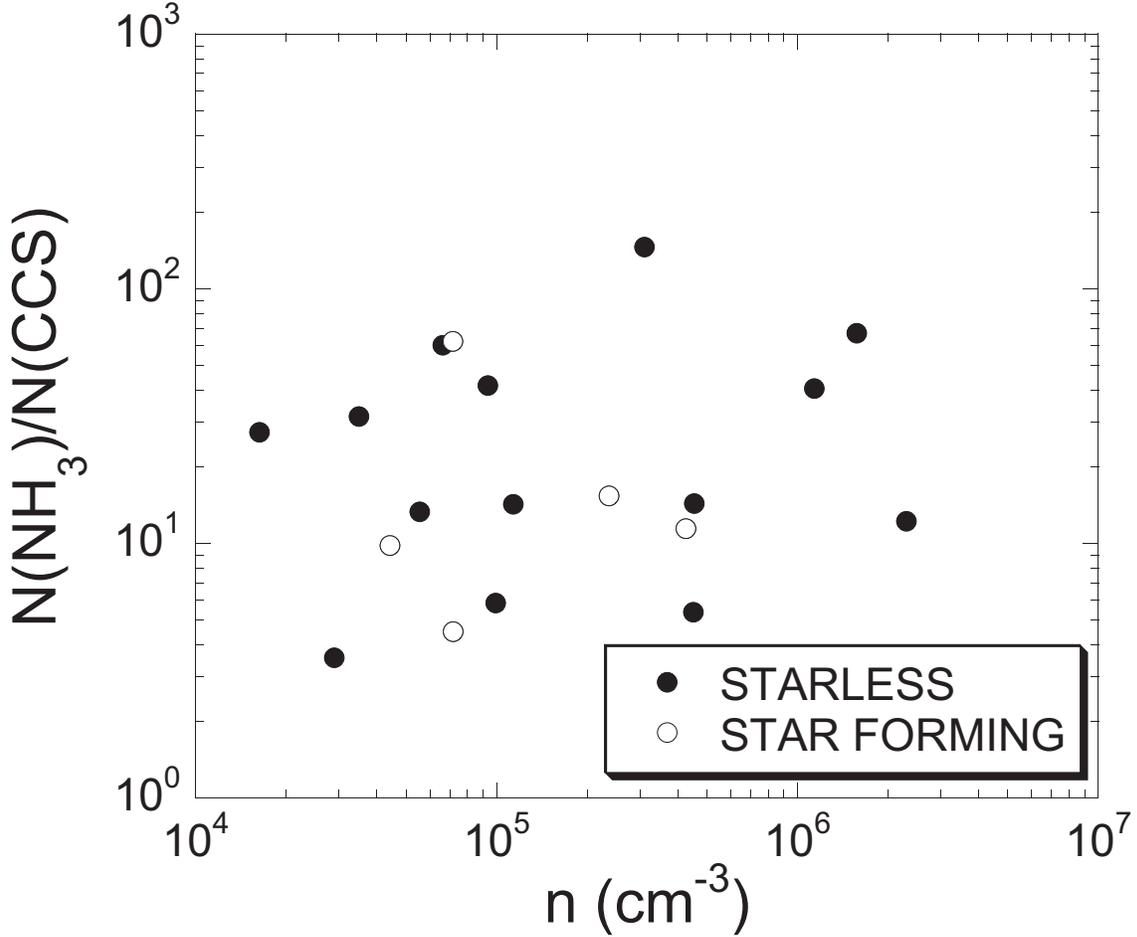}
  \end{center}
  \caption{$N$(NH$_3$)/$N$(CCS) column density ratio 
is plotted against the core average density from
CS observations \citep{tat93}.
The CCS excitation temperature is assumed to be equal to $T_{rot}$.

}\label{fig:figure11}
\end{figure}

\begin{figure}
  \begin{center}
    \FigureFile(150mm,150mm){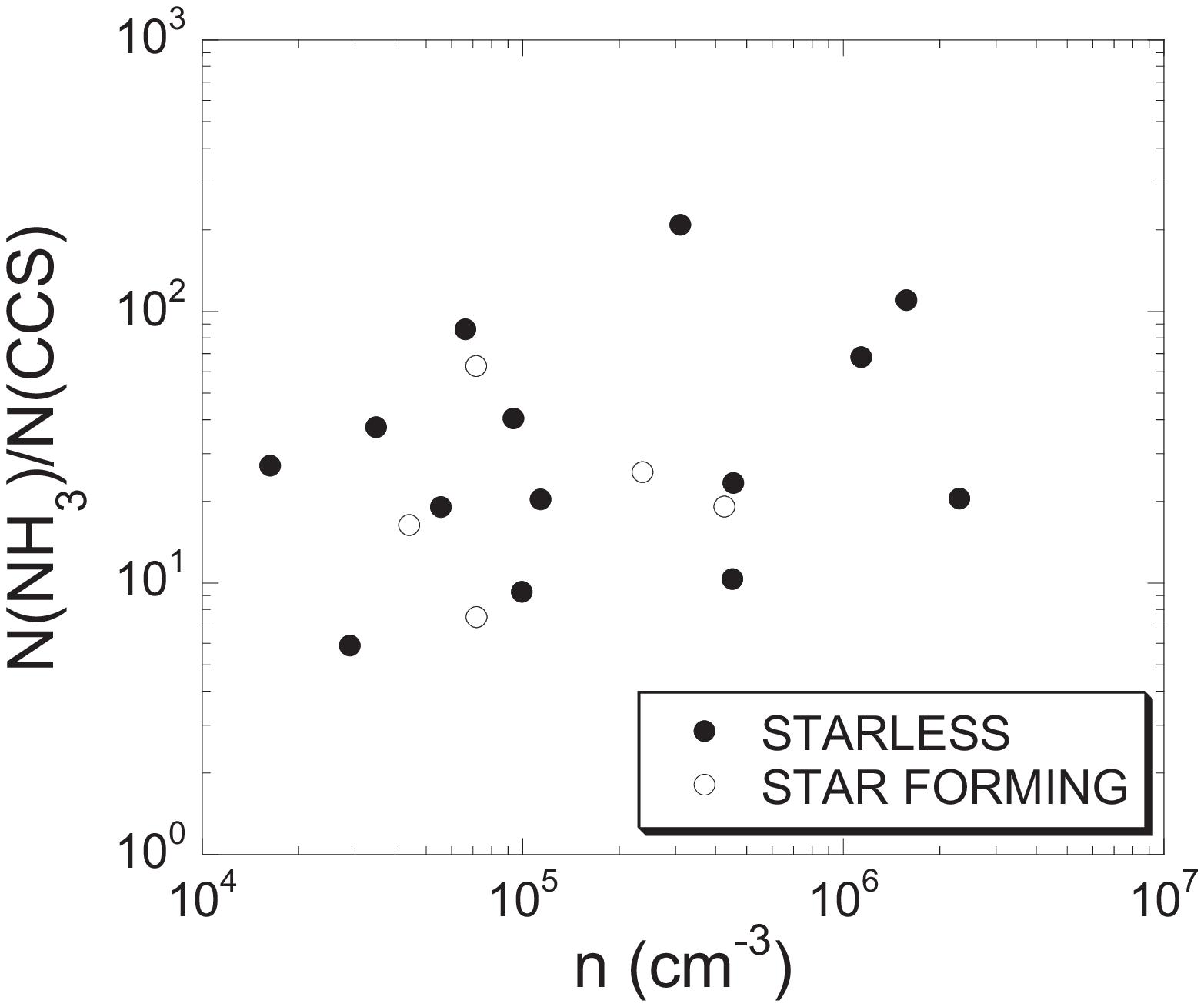}
  \end{center}
  \caption{The same as Figure \ref{fig:figure11} 
but for the CCS excitation temperature equal to $T_{rot}$/2.
}\label{fig:figure12}
\end{figure}

Figures \ref{fig:figure11} and \ref{fig:figure12} show $N$(NH$_3$)/$N$(CCS)
against the core average density $n$, with different CCS excitation temperature
assumptions.
$N$(NH$_3$)/$N$(CCS) and the average density $n$ show no correlation or
a positive, weak
correlation.
A weak positive correlation, if true,
may mean 
that evolved cores tends to have high-density gas 
due to physical evolution and  
high $N$(NH$_3$)/$N$(CCS) due to chemical evolution,
or that cores having high-density gas intrinsically
tends to have higher $N$(NH$_3$)/$N$(CCS) due to faster 
chemical evolution, if we assume $N$(NH$_3$)/$N$(CCS) represents
the chemical evolution as in dark cloud cores.

\begin{figure}
  \begin{center}
    \FigureFile(150mm,150mm){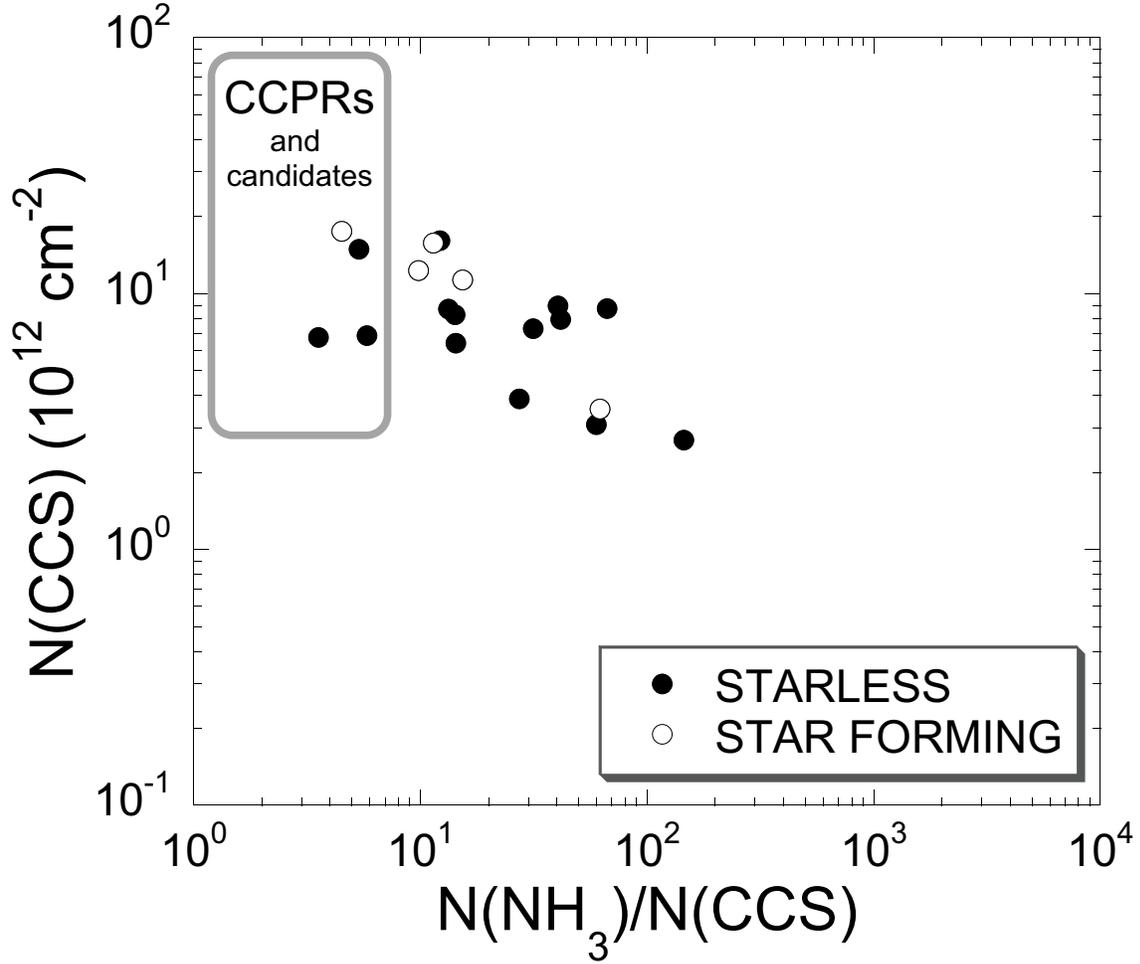}
  \end{center}
  \caption{The $N$(CCS) column density is plotted against
the $N$(NH$_3$)/$N$(CCS) column density ratio.
The CCS excitation temperature is assumed to be equal to 
T$_{rot}$.
The distribution of 
``Carbon-Chain-Producing Regions (CCPRs)'' 
and their candidates defined in
\citet{hir09} is roughly shown as a box.
}\label{fig:figure13}
\end{figure}

\begin{figure}
  \begin{center}
    \FigureFile(150mm,150mm){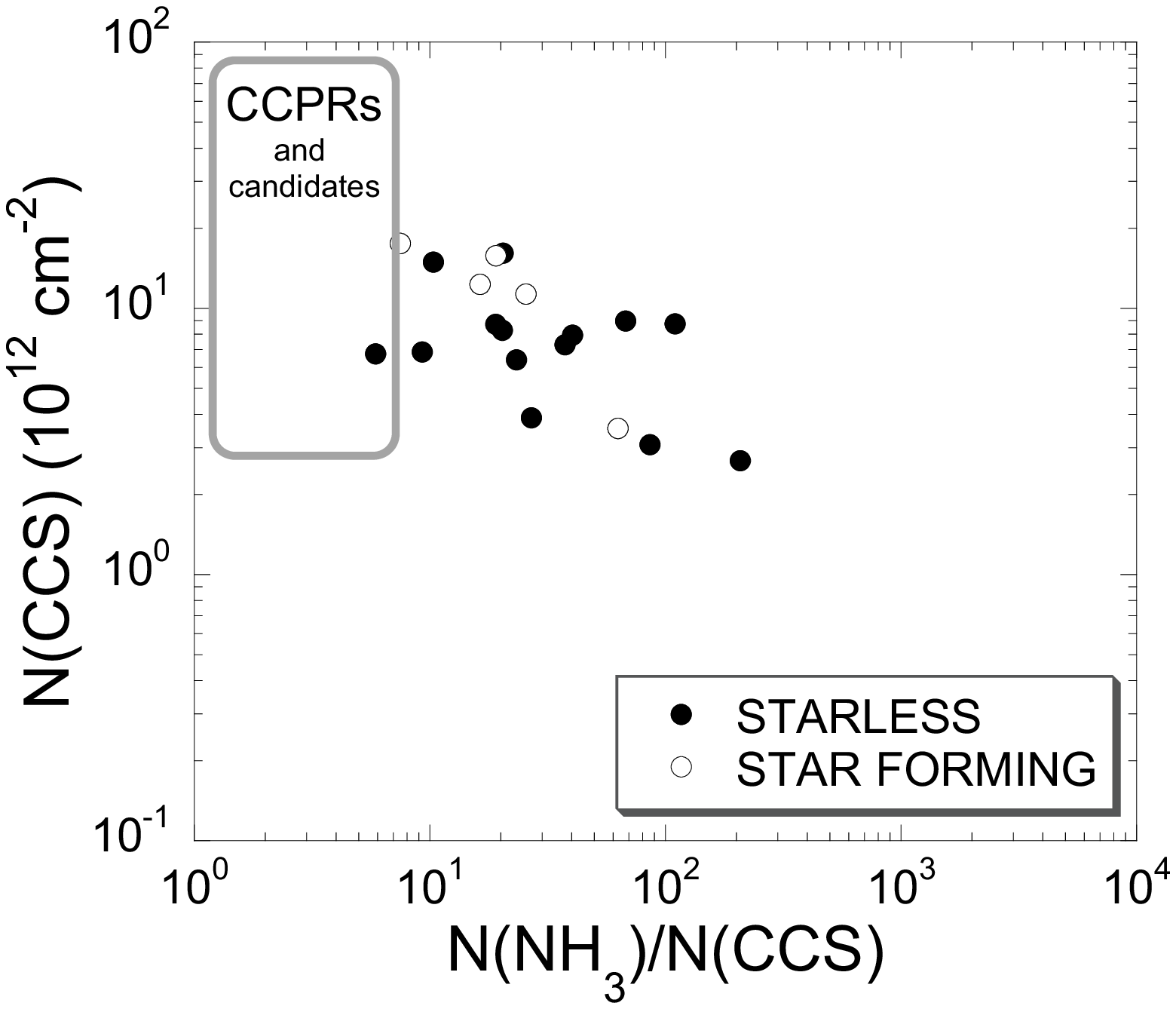}
  \end{center}
  \caption{The same as Figure \ref{fig:figure13} 
but for the CCS excitation temperature equal to $T_{rot}$/2.
}\label{fig:figure14}
\end{figure}

\begin{figure}
  \begin{center}
    \FigureFile(150mm,150mm){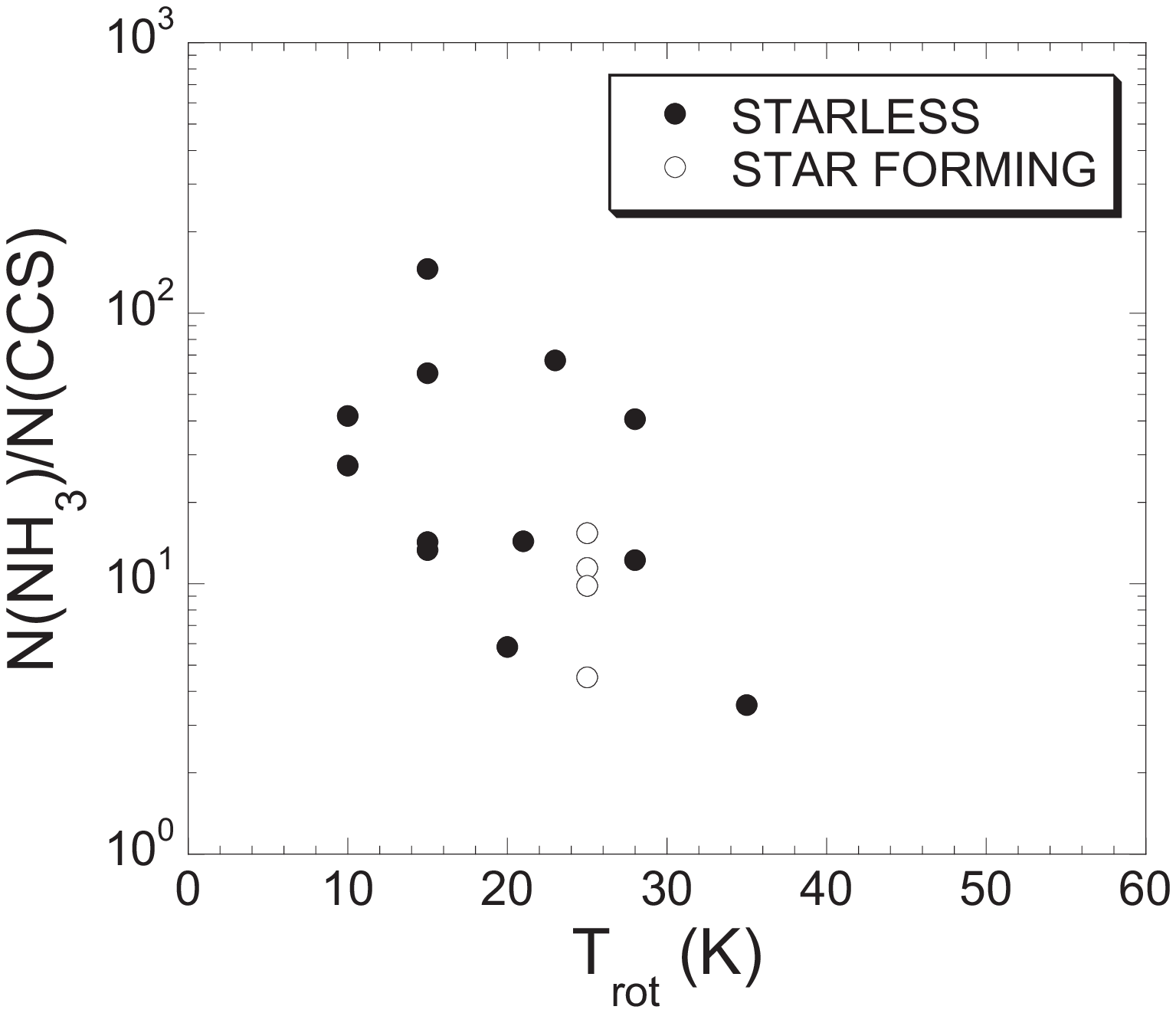}
  \end{center}
  \caption{$N$(NH$_3$)/$N$(CCS) is plotted against $T_{rot}$ for 
the CCS excitation temperature equal to $T_{rot}$.
}\label{fig:figure15}
\end{figure}

\citet{hir09} compared the $N$(NH$_3$)/$N$(CCS) column 
density ratio
among various star forming regions.
Figures \ref{fig:figure13} and \ref{fig:figure14} show $N$(CCS) against 
$N$(NH$_3$)/$N$(CCS)
in the Orion A GMC
to compare with diagrams in \citet{hir09}.
$N$(NH$_3$)/$N$(CCS) ranges
from $\sim$3 to $\sim$100 in Orion A GMC,
from $\sim$10 to $\sim$1000 in Perseus, and
from $\sim$20 to $\sim$1000 in Ophiuchus. 
On the other hand, it ranges
from 2 to 400 in Taurus.
Taurus have cores with very low $N$(NH$_3$)/$N$(CCS) ($\sim$ 2$-$7), 
which corresponds to ``Carbon-Chain-Producing Regions (CCPRs)'' 
\citep{hir09}.
Cores in Perseus and Ophiuchus do not
show very low $N$(NH$_3$)/$N$(CCS).
Figure \ref{fig:figure14}
does not show very low $N$(NH$_3$)/$N$(CCS) in Orion, but
Figure \ref{fig:figure13} shows very low $N$(NH$_3$)/$N$(CCS).
Note that one star-forming core in Orion shows very low $N$(NH$_3$)/$N$(CCS)
in Figure \ref{fig:figure13}.
Orion GMC cores seem not to have high $N$(NH$_3$)/$N$(CCS) ratios
compared with Perseus and Ophiuchus.
It is known that an N-bearing molecule
emitting region is more compact than a CCS emitting region 
in L1544 in Taurus (\cite{aik01}
and references therein).
A possibility is that NH$_3$,
which may have emitting reqions smaller than CCS also in Orion, 
will be beam diluted 
in Orion, which is three times more 
distant than other regions.
The telescope beam size is  $\sim$ 40$\arcsec$ in both of the 
NH$_3$ and 
CCS observations.
We need to check this possibility with higher angular resolution
observations.
Regarding the $N$(NH$_3$)/$N$(CCS)$-$$N$(CCS) diagram,
we should use $N$(CCS)/$N$(H$_2$) instead of $N$(CCS) as the ordinate
to compare with the theoretical calculations of \citet{suz92},
as explained in \citet{hir09}.
Orion GMC cores have higher column densities than Taurus cores.
The average column density of cores derived from N$_2$H$^+$ is
6.7$\times$10$^{22}$ cm$^{-2}$ \citep{tat08} and 
1.7$\times$10$^{22}$ cm$^{-2}$ \citep{tat04} in
the Orion A GMC cores and Taurus cores, respectively.
Here, for Taurus, we used the column density derived from the 
spatially-smoothed
spectrum, which compensates for the difference in linear resolution
due to a factor of 3 difference in distance.
The difference in the average column density 
is a factor of 4,
and we should take into account this ``vertical shift'' 
to compare the $N$(NH$_3$)/$N$(CCS)-$N$(CCS) diagram.
Next, we investigate the origin of CCS.
\citet{ike99} observed the coexistence of the neutral carbon atom and CO
over the Orion A GMC, and suggested that the origin of the neutral carbon atom
might be the photodissociation region due to the cloud clumpiness under
UV irradiation.
We wonder whether CCS in the Orion GMC is related with photodissociation.
CCS was not detected toward cores TUKH034 and TUKH036 
located along the Orion Bar,
which is an archetypal photodissociation region (PDR).
Then, it is not likely that photodissociation in a clumpy cloud
explains CCS in the Orion A GMC.
It is suggested that CCS exists also in evolved molecular gas
as secondary late-stage peak due to CO depletion
(\cite{zli02} and references therein). 
On the other hand,
CO depletion will be less effective in warm molecular gas in Orion.
The gas kinetic temperature of Orion cores ranges from $\sim$ 10 K to $>$ 50 K \citep{wil99}.
Figures \ref{fig:figure15} shows $N$(NH$_3$)/$N$(CCS) against $T_{rot}$.
Warmer $T_{rot}$ cores tend to have low $N$(NH$_3$)/$N$(CCS).
The variation of the NH$_3$ rotation temperatrure $T_{rot}$ toward cores
is shown in \citet{wil99} (TUKH core number
is in declination order).
Very low $N$(NH$_3$)/$N$(CCS) cores in Orion A GMC are warm,
and it seems that they
have physical properties very different from CCPRs in nearby dark cloud cores.
It is not clear whether chemical models can explain CCS abundance
at warm ($T_{rot}$ = 20$-$35 K, $T_{kin}$ = 20$-$55 K) 
Orion cores qualitatively.
We wonder 
why star-forming cores do not tend to have higher $N$(NH$_3$)/$N$(CCS) clearly,
as expected from chemical evolution in dark cloud cores.
In Taurus the displacement of CCS and NH$_3$ is clear because
there are CCPRs.
On the other hand, in Perseus, Ophiuchus, and Orion, 
the displacement is not
very clear.
This may be partly because the range of $N$(NH$_3$)/$N$(CCS)
is narrower than that in Taurus, 
partly because
CCS may exist also in evolved cores
and possibly because the observational errors are too large to see the
separation even if it exists.
Furthermore, we should point out that identification of 
young stellar objects
in Orion
are incomplete because of the diffuse infrared emission of the Orion Nebula
on the Spitzer MIPS image and because of the limited spatial resolution
of the CS observations of $\sim$ 40$\arcsec$ \citep{tat93}.

We found two very interesting Orion cores. 
TUKH122, whose CCS spectrum is shown in Figure \ref{fig:figure2}, 
is the most intense core in CCS, which is strong 
also in NH$_3$.  This core has a substantial amount 
of both CCS and N-bearing molecule. TUKH083,
whose CCS spectrum is shown in Figure \ref{fig:figure1}, is the core with the 
lowest $N$(NH$_3$)/$N$(CCS) column density ratio.
This core has a substantial amount of th early-type molecule CCS.
These two cores look starless through the visual 
inspection of Spitzer 24 micron images. The core mass is 
50-60 solar masses from the CS observations by
\citet{tat93}, and these cores will be very 
good candidates for future cluster formation.

\subsection{Deuterium Fractionation}

\begin{figure}
  \begin{center}
    \FigureFile(150mm,150mm){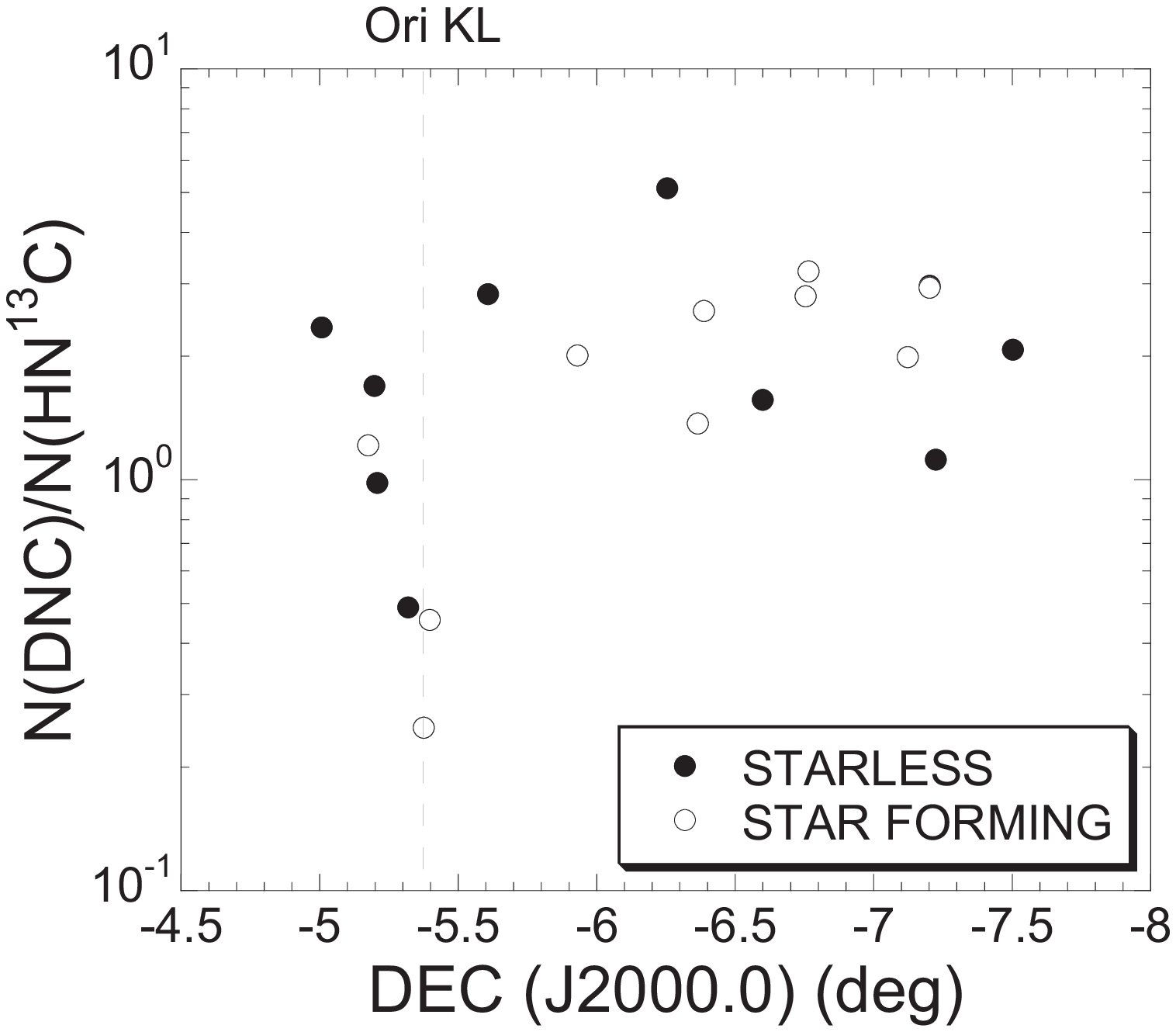}
  \end{center}
  \caption{$N$(DNC)/$N$(HN$^{13}$C) column density ratio 
is plotted against the declination.

}\label{fig:figure16}
\end{figure}

\begin{figure}
  \begin{center}
    \FigureFile(150mm,150mm){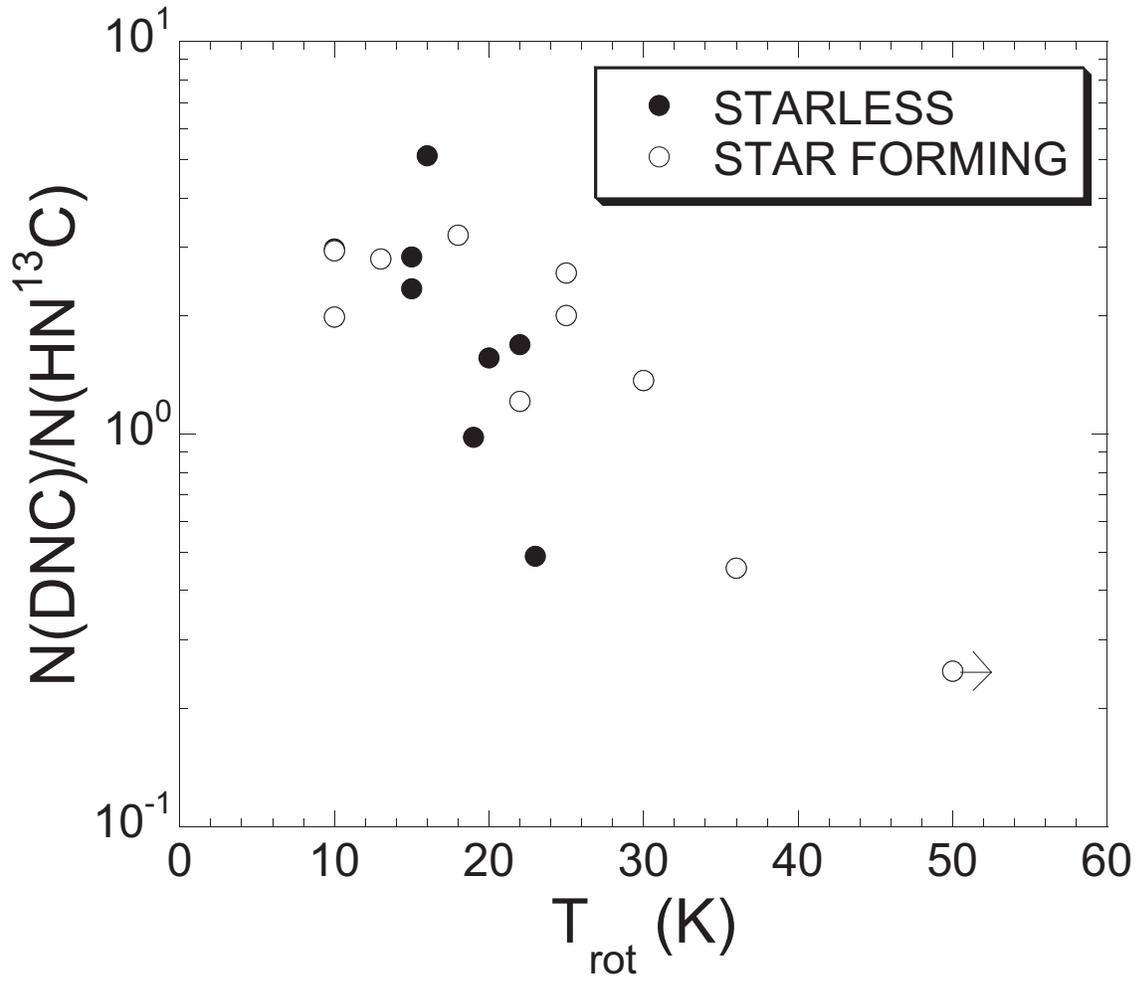}
  \end{center}
  \caption{$N$(DNC)/$N$(HN$^{13}$C) is plotted against the NH$_3$ 
rotation temperature \citep{wil99}.  The lower limit to 
the rotation temperature is shown for Orion KL, which has the lowest 
$N$(DNC)/$N$(HN$^{13}$C).

}\label{fig:figure17}
\end{figure}

\begin{figure}
  \begin{center}
    \FigureFile(150mm,150mm){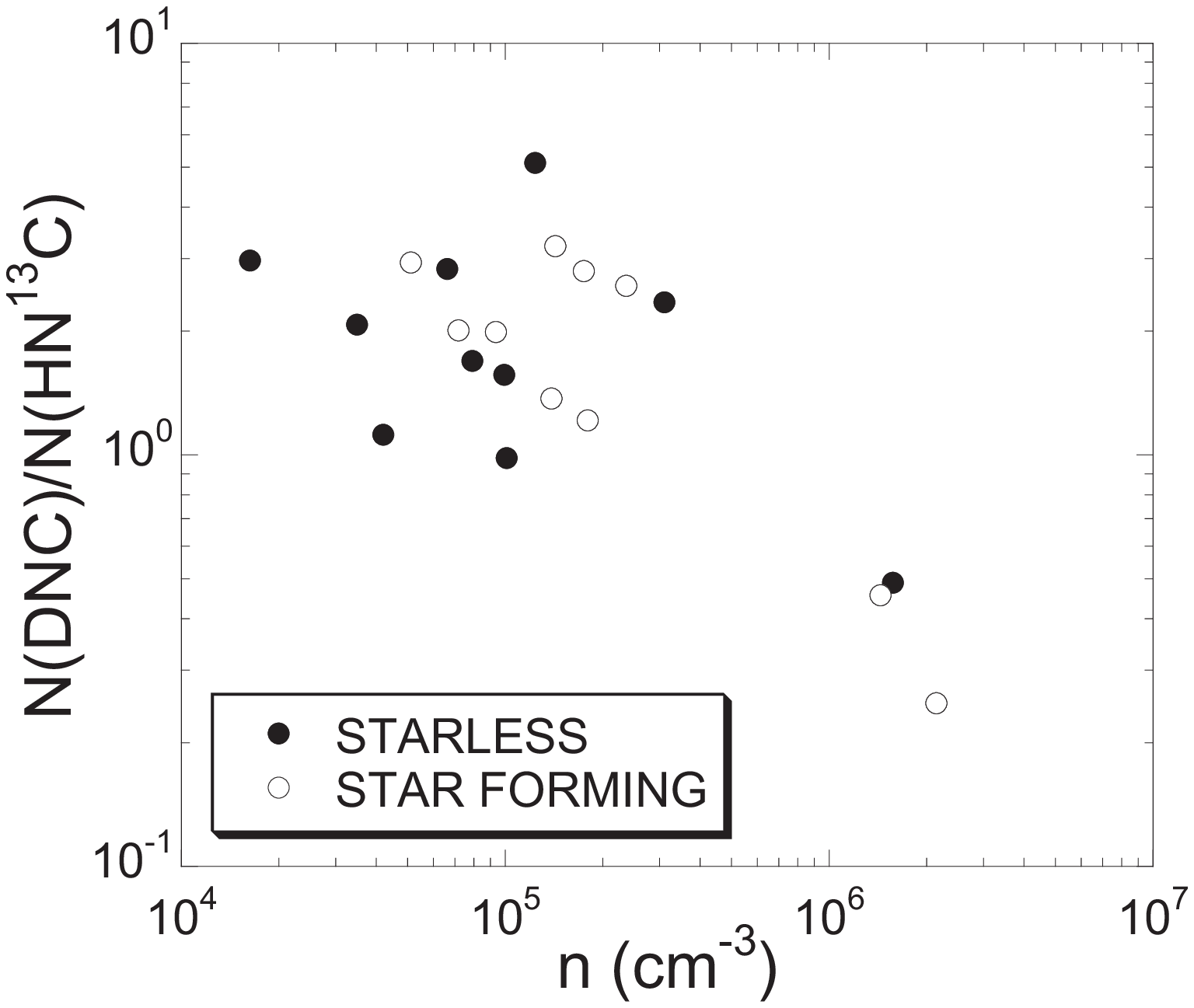}
  \end{center}
  \caption{$N$(DNC)/$N$(HN$^{13}$C) is plotted against the core average
density from CS observations \citep{tat93}.
}\label{fig:figure18}
\end{figure}

We show the column density ratio of DNC to HN$^{13}$C
to see
the deuterium fractionation.
Figures \ref{fig:figure16} plots $N$(DNC)/$N$(HN$^{13}$C) 
column density ratio against the declination.
Because the LTE column density ratio from molecular 
lines at similar frequencies
depends on $T_{ex}$ only very weakly, we show the result
only for $T_{ex}$ = $T_{rot}$.
Around Orion KL, the ratio has the minimum values.
Figures \ref{fig:figure17} plots $N$(DNC)/$N$(HN$^{13}$C) against 
the NH$_3$ rotation temperature \citep{wil99}.
The ratio is lower for warmer cores.  This indicates that the deuterium 
enhancement is less efficient in warm cores.  Similar results have been shown previously by, e.g., \citet{sne79}, \citet{woo87}, and \citet{sch92}. 
We conclude that the tendency seen in $N$(DNC)/$N$(HN$^{13}$C)
is due to variation in the gas temperature:
D/H decreases with increasing gas kinetic temperature.
There are two possibilities for this:
dependence of exchange reactions between HD and molecular ion
on temperature
\citep{sne79} or dependence of CO depletion on temperature.
At low temperatures, the depletion of molecules becomes efficient,
the H$_2$D$^+$/H$_3^+$ abundance ratio increases,
and then molecular D/H ratios increase
\citep{rob00}.
It is found that the deuterium fractionation is a tracer 
of the chemical age of dark cloud cores \citep{hir06}.
However, the variation observed in the present study
is due to variation of gas temperature rather than
core evolution, because Orion cores do not have CCPRs.
Figures \ref{fig:figure18} plots $N$(DNC)/$N$(HN$^{13}$C) against 
the core average density.
It seems that high-density cores tend to show low $N$(DNC)/$N$(HN$^{13}$C).
Depletion will be more serious in high-density cores
at cold temperatures, and therefore the above result cannot be explained
in terms of different depletion degrees due to different density.
We conclude that different $N$(DNC)/$N$(HN$^{13}$C) is due to
difference in gas temperature, not due to difference in density.

\section{Summary} 

We have detected CCS in the Orion A GMC for the first time.
CCS-detected cores are not localized in the Orion cloud and
widely distributed. 
The CCS peak intensity tends to be
high in southern cores.
The HC$_3$N peak intensity of cores tends to be high
toward the south, while there are also HC$_3$N intense cores
near Orion KL.  The core associated with Orion KL show
intense wing emission, and star formation activity 
near Orion KL seems
to enhance the HC$_3$N intensity.
The NH$_3$/CCS column density ratio is higher in the north
and
south regions and lower in the middle region,
suggesting the chemical variation
along the Orion A GMC filament.
This ratio is a good tracer of chemical evolution in nearby dark cloud cores,
but seems to be affected by core gas temperature in the Orion A GMC.
This ratio tends to be low for warmer cores.
The DNC to HN$^{13}$C
column density ratio
is generally similar to that observed in dark cloud cores,
but becomes lower around Orion KL.
Variation in the DNC to HN$^{13}$C column density ratio
seems to reflect the temperature variation among cores.

\bigskip

K. T. thanks Jeong-Eun Lee and Satoshi Yamamoto
for discussion.


\begin{longtable}{lrclrclrclrcl}
  \caption{Orion A Core Line Data}\label{tab:LTsample}
  \hline              
 & & CCS& & &HC$_3$N& & &HN$^{13}$C& & &DNC& \\
\hline
TUKH & $T_A^*$&$v_{LSR}$&$\Delta v$&$T_A^*$&$v_{LSR}$&$\Delta v$&$T_A^*$&
$v_{LSR}$&$\Delta v$&$T_A^*$&$v_{LSR}$&$\Delta v$\\
\hline
 &K&km s$^{-1}$&km s$^{-1}$&K&km s$^{-1}$&km s$^{-1}$&K&km s$^{-1}$&
km s$^{-1}$&K&km s$^{-1}$&km s$^{-1}$\\
  \hline 
\endhead
  \hline
\endfoot
  \hline
\endlastfoot
  \hline
001&...&...&...&...&...&...&$<$0.09&...&...&$<$0.13&...&...\\
002&...&...&...&...&...&...&$<$0.08&...&...&$<$0.14&...&...\\
003&0.22&10.41&0.45&0.74&11.15&1.01&0.37&11.18&0.89&0.66&11.21&1.15\\
004&...&...&...&...&...&...&$<$0.09&...&...&$<$0.13&...&...\\
006&...&...&...&...&...&...&$<$0.09&...&...&$<$0.14&...&...\\
007&$<$0.09&...&...&$<$0.18&...&...&$<$0.05&...&...&0.29&11.43&0.73\\
008&...&...&...&...&...&...&$<$0.08&...&...&$<$0.13&...&...\\
009&...&...&...&...&...&...&$<$0.09&...&...&$<$0.14&...&...\\
010&$<$0.11&...&...&$<$0.17&...&...&$<$0.04&...&...&$<$0.06&...&...\\
011&$<$0.11&...&...&1.29&11.34&1.33&0.26&11.1&1.21&0.17&10.93&2.22\\
013&$<$0.11&...&...&0.43&10.74&1.03&0.15&10.59&0.97&0.2&10.31&1.18\\
014&...&...&...&...&...&...&$<$0.08&...&...&$<$0.13&...&...\\
015&$<$0.09&...&...&0.43&11.02&0.93&0.21&10.97&0.75&0.27&10.88&0.58\\
016&...&...&...&...&...&...&$<$0.07&...&...&$<$0.13&...&...\\
017&...&...&...&...&...&...&$<$0.08&...&...&$<$0.11&...&...\\
021&0.23&9.32&0.98&2.55&9.59&1.77&0.78&9.49&1.48&0.32&9.68&1.79\\
023&...&...&...&...&...&...&$<$0.07&...&...&$<$0.11&...&...\\
025&$<$0.10&...&...&0.31&6.45&0.69&$<$0.04&...&...&$<$0.06&...&...\\
026&0.25&10.09&0.80&2.91&10.01&1.38&0.36&9.95&1.22&$<$0.07&...&...\\
028&...&...&...&...&...&...&$<$0.08&...&...&0.19&8.2&1.40\\
029&$<$0.11&...&...&1.92&8.33&6.29&0.21&8.73&3.86&0.15&7.86&1.27\\
030&...&...&...&...&...&...&$<$0.08&...&...&$<$0.13&...&...\\
031&$<$0.08&...&...&1.23&6.80&3.01&0.24&6.72&2.64&0.16&6.70&1.79\\
032&$<$0.11&...&...&0.67&11.04&1.2&$<$0.08&...&...&$<$0.12&...&...\\
033&0.13&8.48&1.34&1.13&8.73&1.96&0.16&8.52&1.78&$<$0.07&...&...\\
034&$<$0.10&...&...&$<$0.15&...&...&$<$0.06&...&...&$<$0.09&...&...\\
036&$<$0.14&...&...&0.33&8.16&1.99&$<$0.08&...&...&$<$0.14&...&...\\
037&0.21&8.88&1.70&1.24&8.79&1.96&0.23&8.50&2.25&$<$0.07&...&...\\
039&0.10&8.64&3.79&2.55&9.93&1.34&0.43&9.85&0.91&$<$0.07&...&...\\
040&0.25&11.34&1.43&0.92&11.18&0.91&$<$0.09&...&...&$<$0.12&...&...\\
042&$<$0.10&...&...&$<$0.15&...&...&0.14&6.79&0.73&$<$0.13&...&...\\
044&$<$0.12&...&...&$<$0.14&...&...&0.17&7.05&0.65&$<$0.13&...&...\\
046&$<$0.13&...&...&$<$0.14&...&...&$<$0.08&...&...&0.3&7.78&0.63\\
047&0.19&8.51&0.59&0.35&8.12&1.74&0.35&8.14&0.66&0.75&8.17&0.85\\
048&$<$0.08&...&...&$<$0.11&...&...&$<$0.05&...&...&$<$0.07&...&...\\
049&0.24&7.53&1.26&$<$0.16&...&...&$<$0.10&...&...&$<$0.15&...&...\\
055&$<$0.14&...&...&$<$0.16&...&...&$<$0.09&...&...&0.31&10.18&0.99\\
056&0.22&8.09&1.97&0.26&7.37&0.90&0.31&7.49&0.44&0.29&7.43&0.93\\
057&0.22&9.03&1.43&$<$0.15&...&...&$<$0.09&...&...&$<$0.14&...&...\\
059&0.21&8.24&1.42&$<$0.14&...&...&$<$0.09&...&...&$<$0.14&...&...\\
060&$<$0.16&...&...&$<$0.17&...&...&$<$0.09&...&...&$<$0.15&...&...\\
063&$<$0.16&...&...&$<$0.15&...&...&$<$0.09&...&...&$<$0.15&...&...\\
065&$<$0.08&...&...&0.31&7.65&0.68&0.16&7.37&0.32&0.25&7.46&1.03\\
066&$<$0.12&...&...&$<$0.17&...&...&$<$0.09&&&$<$0.14&...&...\\
067&$<$0.11&...&...&0.35&6.86&2.94&0.26&7.16&1.15&0.26&7.17&1.55\\
068&...&...&...&...&...&...&$<$0.14&&&$<$0.21&...&...\\
069&0.18&6.98&1.57&0.40&6.75&2.25&0.17&6.55&1.8&0.54&7.09&1.40\\
071&...&...&...&...&...&...&$<$0.14&...&...&$<$0.22&...&...\\
072&...&...&...&...&...&...&$<$0.15&...&...&0.32&8.3&0.88\\
073&...&...&...&...&...&...&$<$0.15&...&...&$<$0.19&...&...\\
079&...&...&...&...&...&...&$<$0.13&...&...&$<$0.19&...&...\\
081&$<$0.11&...&...&$<$0.16&...&...&$<$0.06&...&...&$<$0.08&...&...\\
082&...&...&...&...&...&...&$<$0.15&...&...&$<$0.20&...&...\\
083&0.35&7.43&0.37&0.16&8.44&3.50&$<$0.05&...&...&$<$0.08&...&...\\
084&...&...&...&...&...&...&$<$0.15&...&...&$<$0.22&...&...\\
085&$<$0.20&...&...&$<$0.28&...&...&$<$0.09&...&...&$<$0.11&...&...\\
087&$<$0.09&...&...&$<$0.14&...&...&$<$0.05&...&...&0.11&7.07&1.43\\
088&0.31&6.46&0.65&0.54&6.67&1.01&0.14&6.56&0.8&0.19&6.49&0.95\\
090&$<$0.17&...&...&$<$0.26&...&...&$<$0.08&...&...&$<$0.12&...&...\\
091&$<$0.12&...&...&0.38&8.90&1.41&0.14&8.89&1.65&0.35&8.88&1.81\\
092&$<$0.12&...&...&0.27&10.30&1.65&0.14&9.69&1.56&0.22&9.95&3.22\\
093&$<$0.19&...&...&0.24&8.71&2.41&$<$0.08&...&...&$<$0.12&...&...\\
095&$<$0.22&...&...&0.36&4.03&1.62&$<$0.08&...&...&$<$0.11&...&...\\
096&$<$0.19&...&...&$<$0.25&...&...&$<$0.09&...&...&0.3&6.04&0.73\\
097&0.33&5.64&1.09&0.73&5.91&0.9&0.27&5.6&1.04&0.47&5.54&1.16\\
099&$<$0.09&...&...&$<$0.13&...&...&$<$0.05&...&...&$<$0.07&...&...\\
100&$<$0.19&...&...&$<$0.29&...&...&$<$0.12&...&...&$<$0.16&...&...\\
103&$<$0.10&...&...&$<$0.14&...&...&0.12&5.13&1.06&0.36&5.32&1.01\\
104&0.26&2.97&0.70&0.86&3.02&0.82&0.28&2.94&0.50&0.38&3.08&1.07\\
105&0.19&6.15&0.88&0.28&6.07&0.64&0.17&6.40&1.00&0.19&6.32&0.97\\
106&$<$0.17&...&...&$<$0.28&...&...&$<$0.11&...&...&$<$0.15&...&...\\
107&$<$0.19&...&...&$<$0.30&...&...&$<$0.10&...&...&$<$0.15&...&...\\
111&$<$0.18&...&...&$<$0.27&...&...&$<$0.10&...&...&0.27&4.58&0.70\\
112&$<$0.19&...&...&$<$0.31&...&...&$<$0.11&...&...&$<$0.15&...&...\\
114&$<$0.16&...&...&$<$0.28&...&...&$<$0.11&...&...&$<$0.14&...&...\\
116&$<$0.10&...&...&$<$0.15&...&...&$<$0.05&...&...&$<$0.07&...&...\\
117&0.16&4.46&1.04&0.48&4.33&0.66&$<$0.05&...&...&$<$0.07&...&...\\
121&$<$0.19&...&...&$<$0.19&...&...&$<$0.12&...&...&0.54&3.86&0.31\\
122&0.47&3.86&0.55&1.70&3.84&0.78&0.42&3.74&0.72&0.60&3.77&1.03\\
123&$<$0.17&...&...&0.48&4.55&0.86&$<$0.11&...&...&0.32&4.87&1.06\\
125&$<$0.20&...&...&0.53&3.84&0.86&$<$0.12&...&...&$<$0.20&...&...\\
\end{longtable}



\end{document}